\documentclass[12pt]{book}
\usepackage{psfig}
\usepackage{rotating}

\def\wig#1{\mathrel{\hbox{\hbox to 0pt{%
          \lower.5ex\hbox{$\sim$}\hss}\raise.4ex\hbox{$#1$}}}}
\def\Teff{T_{\rm eff}}
\renewcommand{\baselinestretch}{1.1}

\begin{document}
	\thispagestyle{empty}

\centerline{\bf Direct Detection of Galactic Halo Dark Matter}

\medskip

\centerline{B.R.~Oppenheimer, N.C.~Hambly, A.P.~Digby, S.T.~Hodgkin,
D. Saumon\footnote{B.R.~Oppenheimer (bro@astron.berkeley.edu) is
at Astronomy Department, University of California--Berkeley, Berkeley,
CA 94720-3411, USA.  N.C.~Hambly (nch@roe.ac.uk) and A.~P.~Digby
(apd@roe.ac.uk) are at Institute for Astronomy, University of
Edinburgh, Royal Observatory, Blackford Hill, Edinburgh, EH9 3HJ, UK.
S.T.~Hodgkin (sth@ast.cam.ac.uk) is at Institute of Astronomy,
Cambridge University, Madingley Road, Cambridge, CB3 0HA, UK.
D.~Saumon (dsaumon@cactus.phy.vanderbilt.edu) is at Department of
Physics and Astronomy, Vanderbilt University, Nashville, TN 37235
USA.}}

\bigskip

\noindent{\bf The Milky Way Galaxy contains a large, spherical
component which is believed to harbor a substantial amount of unseen
matter.  Recent observations indirectly suggest that as much as half
of this ``dark matter'' may be in the form of old, very cool white
dwarfs, the remnants of an ancient population of stars as old as the
Galaxy itself.  We conducted a survey to find faint, cool
white dwarfs with large space velocities, indicative of their
membership in the Galaxy's spherical halo component.  The survey
reveals a substantial, directly observed population of old white
dwarfs, too faint to be seen in previous surveys.  This newly
discovered population accounts for at least 2\% of the halo dark
matter.  It provides a natural explanation for the indirect
observations, and represents a direct detection of Galactic halo dark
matter.}

\medskip

Dark matter in the spherical halo of the Milky Way Galaxy has been
inferred because the gravitational field due to the known distribution
of luminous matter, primarily stars, cannot explain the observed
rotational characteristics of the Galaxy's spiral disk.  A substantial
portion of this unseen matter may be old, very cool white dwarfs ({\it
1-4}).  A white dwarf is the extremely dense end-state in the
evolution of stars with masses less than about eight times the mass
of the Sun ($M_\odot$).  Once a star becomes a white dwarf, it no
longer produces energy through nuclear fusion, and therefore cools and
fades.  The first four examples of ultracool white dwarfs, whose
temperatures are below 4000 K, were identified only in the past 2
years ({\it 2, 3, 5-7}), principally because they were too faint to
detect in previous surveys.  They have spectral energy distributions
that differ dramatically from those of the hotter white dwarfs,
consistent with white dwarf atmosphere models ({\it 8-10}).  The
difference is due to the formation of H$_2$ molecules in white dwarf
atmospheres with effective temperatures below $\sim 4500$ K.  Although
H$_2$ molecules are symmetric and thus have no dipole moment, in the
high densities of white dwarf atmospheres, collisions between the
molecules are common.  These collisions induce momentary dipole
moments, which produce opacity at wavelengths longer than 0.6 $\mu$m
({\it 11}).  We conducted a survey to search for nearby,
high-proper-motion white dwarfs that could be halo members and might
exhibit this opacity.  Regardless of their spectral appearance, nearby
halo stars can be distinguished from disk stars by their space motion,
because the two populations possess different kinematic properties.

\clearpage
\centerline{\bf Observations}

Our survey used digitized, photographic plates in the $R59F$ and
$B_{\rm J}$ passbands from the SuperCOSMOS Sky Survey ({\it 12-14}).
$R59F$ and $B_{\rm J}$ roughly correspond to optical wavelengths of
0.59 and 0.45 $\mu$m respectively.  We searched for objects with
proper motions, $\mu$, between 0.33 and 10.00 $^{\prime\prime}$
year$^{-1}$ as faint as $R59F = 19.8$, using plates near the South
Galactic Cap (SGC) at three epochs in each field.  The SGC plates
comprise 196 fields ({\it 15}).  The non overlapping area of each
field is 25 square degrees. Hence, the area of our survey is 4900
square degrees, or about 12\% of the sky.  Because of image blending
problems and large halos of scattered light around bright stars, the
actual area surveyed is 4165 square degrees, or 10\%\ of the sky.
This is several times larger than the areas searched in two previous
surveys ({\it 3, 16}).  At the very faint end of our sample, we found
many objects not included in the Luyten catalogs of proper motion
stars ({\it 17}), primarily because Luyten's survey in the Southern
Hemisphere did not attempt to find objects as faint as those we have
found.  These previously undetected stars are the principal members of
our sample of halo white dwarfs.

\begin{figure}
\centerline{\psfig{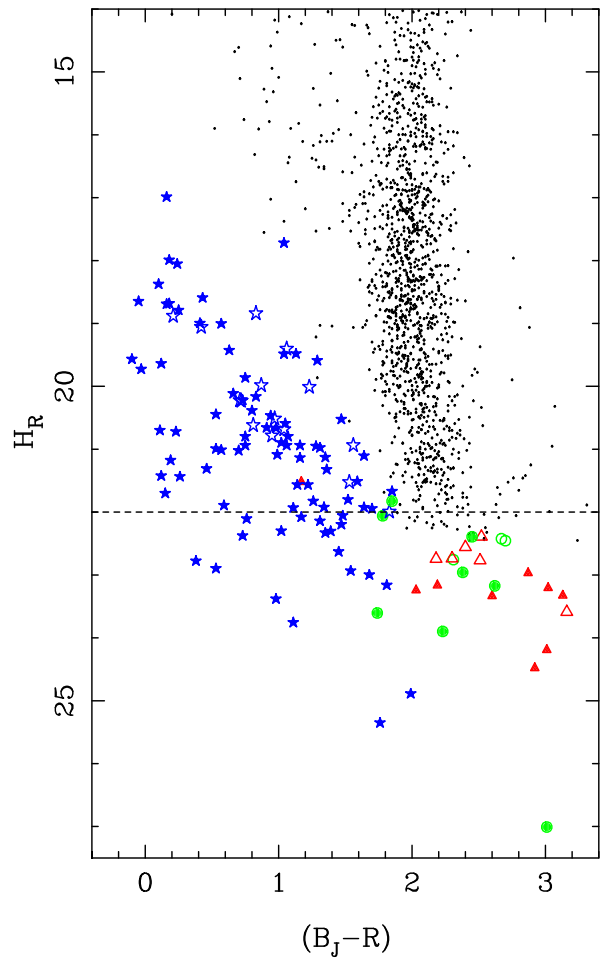}}
\caption{Reduced proper motion diagram.  All low-luminosity,
proper-motion objects in our survey are shown.  The main sequence
extends from the top-center downward.  The subluminous sequence to the
left and below is the white dwarf sequence.  Blue, filled stars
indicate white dwarfs confirmed with our spectroscopy or previously
known to be white dwarfs.  Open stars are objects we have not yet
observed spectroscopically but presume to be white dwarfs on the basis
of their location in this diagram.  Filled, red triangles are M-dwarfs
and filled, green circles are subdwarfs for which we obtained spectra.
Open triangles and open circles are suspected M-type dwarf or
sub-dwarf stars, respectively, which we did not observe.  The dashed
line indicates the value of H$_{\rm R}$ below which the Luyten
catalogs ({\it 17}) contain very few objects.\label{fig:rpm}}
\end{figure}

We used the technique of reduced proper motion (RPM) ({\it 18, 19}) to
identify white dwarfs by their subluminosity in comparison to
main-sequence stars and their high intrinsic space motions.  RPM is
defined as $H_{\rm R} = R59F +$ 5 log $\mu + 5$, and is an estimate
of the absolute magnitude based on the proper motion, $\mu$, in
arcseconds per year.  We selected all 126 objects in the RPM diagram
(Fig.~\ref{fig:rpm}) having either $H_{\rm R}>22.5$ or lying in the
subluminous regions as potential halo white dwarfs.  Sixty-three of
these were previously cataloged ({\it 12, 20}). Of these 63 known,
high-proper-motion stars, 29 had no published spectroscopic
follow-up. This yielded an observing list of 92 stars for follow-up,
including 63 new discoveries.

\begin{figure}
\centerline{\psfig{file=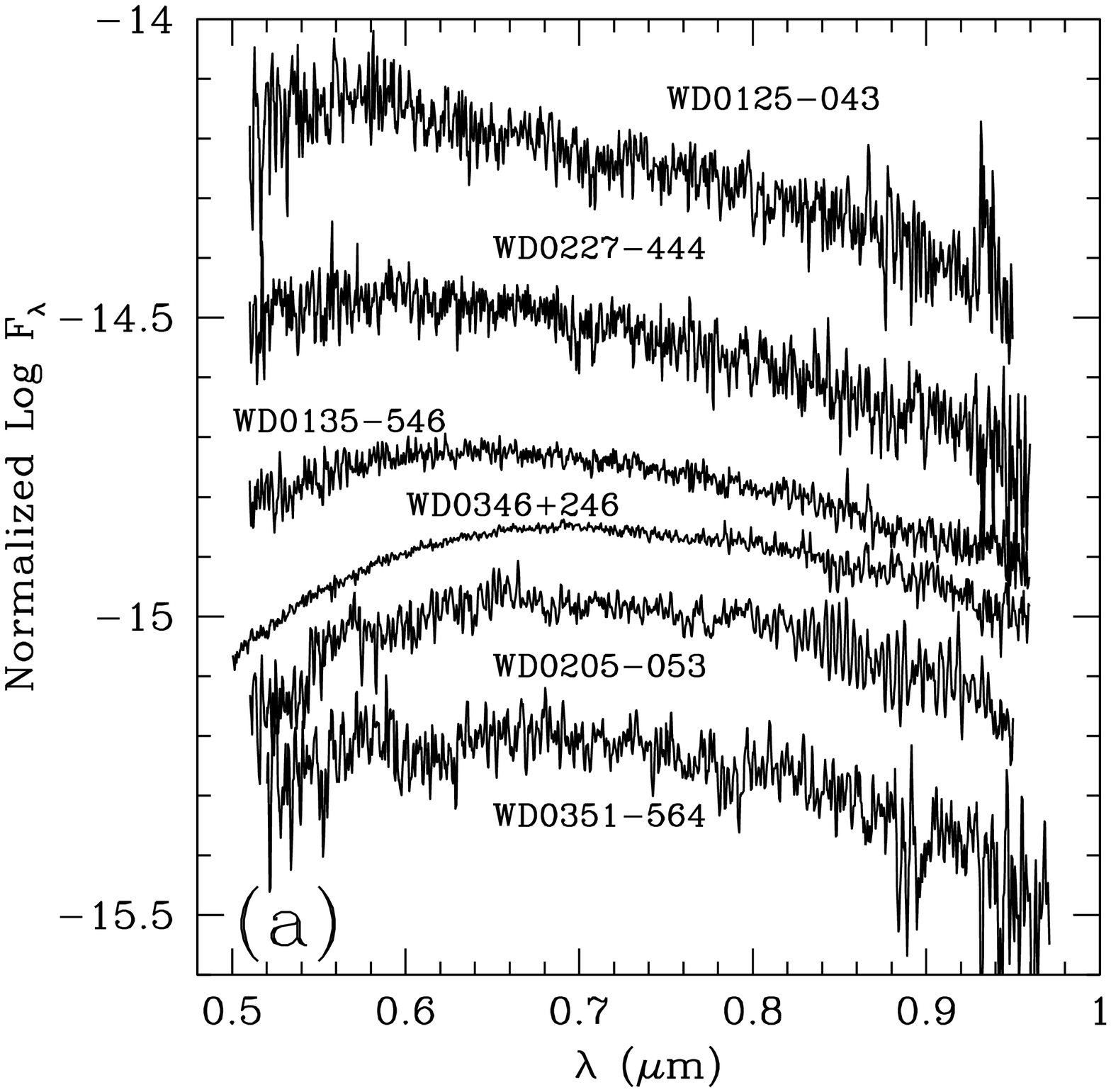,height=3.5in}\psfig{file=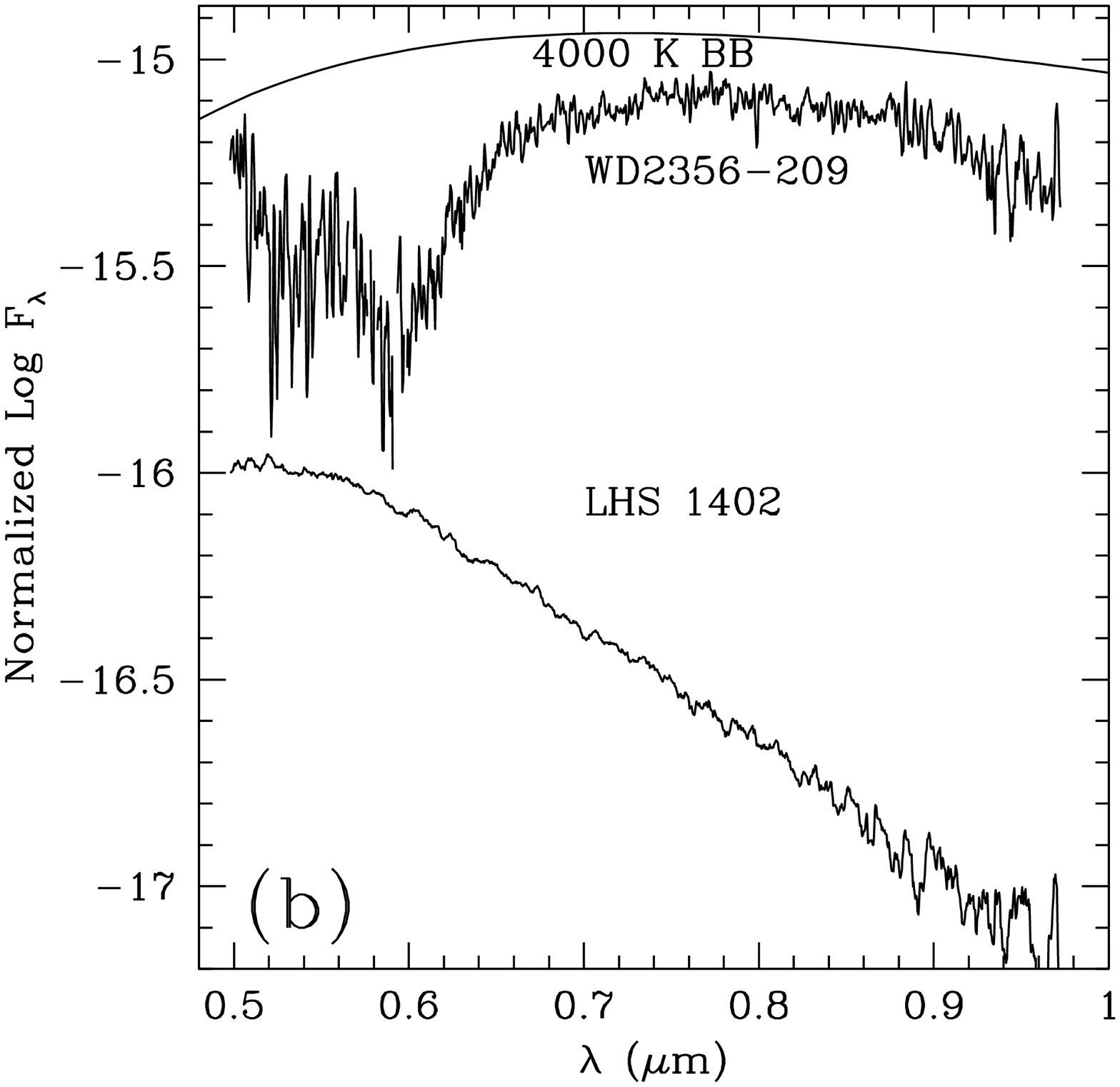,height=3.5in}}

\caption{Selected spectra of new cool white dwarfs.  WD0346$+$246 is
included for comparison.  (a) contains five spectra representative of
the set of 24 cool halo white dwarfs (without H$\alpha$ features)
such that the full range of spectral shapes in this set is presented.
The spectra, normalized and shifted vertically to facilitate
comparison, are ordered according to the sequence in Fig.\
\ref{fig:color}, from top down in order of increasing $B_{\rm J} -$
R59F.  This sequence roughly corresponds to decreasing temperature.
(b) shows the odd spectrum of WD2356$-$209, which has no analogs, the
spectrum of LHS 1402, which seems to be a cooler analog of LHS 3250
and SDSS 1337$+$00, and a 4000 K blackbody spectrum (BB) at the top. Object
names used in Table 1 are indicated.  None of the fine
features are real.\label{fig:sample}}
\end{figure}

Over four nights (11 to 14 October 2000) at the Cerro Tololo
Interamerican Observatory's V.\ M.~Blanco 4-m Telescope, we obtained
spectra in the wavelength range 0.48 to 0.98 $\mu$m for 69 of the 92
candidates (filled symbols in Fig.\ \ref{fig:rpm}).  We inspected the
spectra within minutes of collecting the data, so we were able to
discern by the middle of the second night which types of objects
populated which parts of the RPM diagram (Fig.\ \ref{fig:rpm}).  For
example, as expected, we found only white dwarfs with H$\alpha$
features on the far left of the subluminous sequence.  Dwarf and
subdwarf M--type stars are restricted to the right, which is also the
bottom of the main sequence.  We concentrated our observing resources
on the cool objects, which occupy the lower part of the subluminous
sequence.  Of the 69 candidate objects observed, 16 are M-dwarfs or
M-subdwarfs, 2 are hot He white dwarfs, 13 are white dwarfs with
H$\alpha$ features, and 38 are new cool white dwarfs (Fig.\
\ref{fig:sample}a), a few of which are probably cooler than
WD0346$+$246, the prototypical ultracool white dwarf, which exhibits
the H$_2$ opacity ({\it 2, 7, 21, 22}).

Two of the 38 new cool white dwarfs have unusual spectra (Fig.\
\ref{fig:sample}b).  One, LHS~1402, has a spectrum very similar to
those of the peculiar stars LHS~3250 ({\it 5, 7}) and SDSS~1337+00
({\it 6}), but with a steeper slope toward longer wavelengths,
suggesting a cooler temperature.  The other object, WD2356$-$209,
possesses a bizarre spectrum, incomparable to any other known object.
We reanalyzed the data on this object several times and found no
evidence for residual instrumental effects.  

\smallskip

\centerline{\bf White Dwarfs in the Galactic Halo} 

The 38 cool white dwarfs in our sample show no spectral lines (e.g.,
Fig.\ \ref{fig:sample}a), preventing determination of their radial
velocities.  Nevertheless, we can estimate their space motions from
the tangential velocity, $v_{\rm tan} = \mu d$, where $\mu$ is the
proper motion and $d$ is the distance between each star and Earth.  To
estimate distances, we used a photometric parallax relation derived
from a linear least--squares fit to the cool white dwarf sample of
Bergeron, Ruiz, and Leggett ({\it 23}), supplemented by the ultracool
WD0346+246, yielding the absolute magnitude in the $B_{\rm J}$ filter,
$M_{{\rm B}_{\rm J}}=12.73+2.58(B_{\rm J} - R59F)$.  Our calculations
of the spectra of very cool white dwarfs also show a linear relation
between $M_{{\rm B}_{\rm J}}$ and $B_{\rm J} - R59F$.  The color
turnoff in $B_{\rm J} - R59F$ due to collision-induced absorption by
H$_2$ molecules in the pure hydrogen atmospheres becomes apparent in
$B_{\rm J} - R59F$ only at effective temperatures below 2500$\,$K.
The distances derived by this method are listed in Table 1.  The
scatter of the data around the least-squares fit results in a 20\%
uncertainty in our distance estimates.  For LHS 147 and LHS 542, the
only stars in Table 1 with measured distances ({\it 16, 24}), our
estimates and the measured distances agree within 1 and 20\%,
respectively.

\begin{figure}

\centerline{\psfig{file=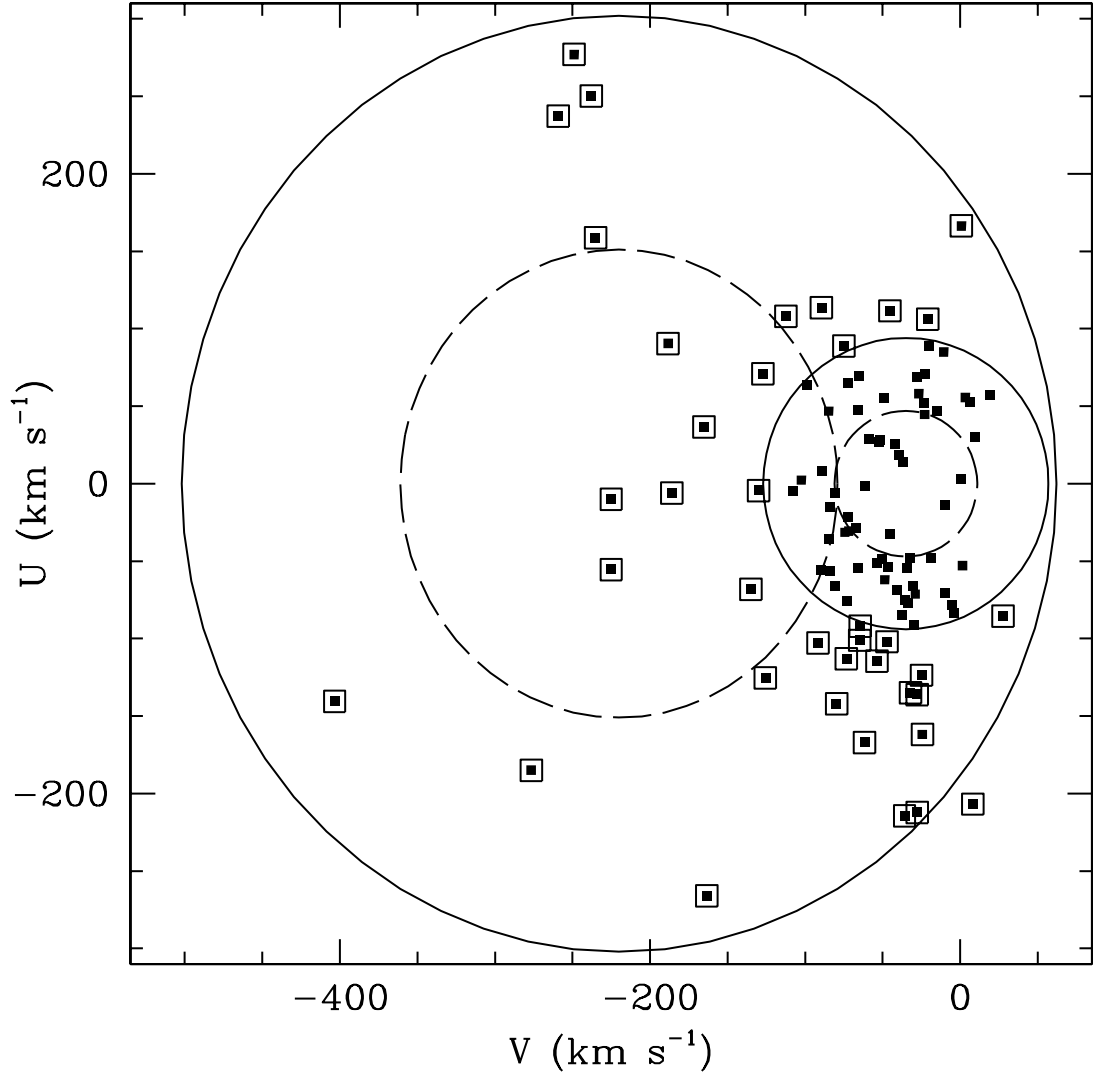,width=6.0in}}

\caption{Velocities of white dwarfs in the survey.  The terms $U$ and
$V$ are the Galactic radial and rotational velocities, respectively,
of the white dwarfs in our survey.  Objects that appear in Table 1 are
surrounded by boxes.  The dashed ellipses indicate the velocity
dispersions of the Galaxy's old disk (right) and halo (left).  The
solid ellipses are the 2$\sigma$ dispersions for these populations.
The ellipses representing the old disk stars, the highest velocity
members of the Galaxy's disk, are centered at $(V, U) = (-35, 0)$,
because this population of stars rotates slightly slower than the
local standard of rest.
\label{fig:velocity}} 
\end{figure}

In order to use the velocities of the stars as an indicator of
membership in the Galactic Halo, we transformed $v_{\rm tan}$ into the
components $U$, $V$, and $W$, in galactic coordinates.  $U$ is radial
away from the Galactic center, $V$ is in the direction of rotation,
whereas $W$ is perpendicular to the Galactic disk.  The calculations
used to derive these velocities take into account the deviation of the
velocity of the Sun with respect to the average velocity of nearby
stars in the Galactic disk.  Because our survey was centered around
the SGC, our observations are most sensitive to the $U$ and $V$
velocities (Fig.~\ref{fig:velocity}).  Therefore, the most reasonable
assumption is that $W$ is zero for all of our stars.  Zero is also the
average value of $W$ for all stars in the Milky Way.  The point where
$U$, $V$, and $W$ are all equal to zero defines the so-called ``local
standard of rest,'' a frame of reference which is rotating in the disk
about the Galactic center at the average rotation velocity of the disk
near the Sun.  Halo objects, because they have an isotropic velocity
distribution with little or no overall rotation, should lag behind the
local standard of rest with a distribution centered near $V = -220$ km
s$^{-1}$, because the local standard of rest orbits the Galaxy at 220
km s$^{-1}$ ({\it 25}).

Our survey contains 38 stars that unambiguously possess halo
kinematics (Fig.\ \ref{fig:velocity}).  We selected a sample of halo
stars by accepting all objects whose velocities exceed the 2$\sigma =
94$ km s$^{-1}$ velocity dispersion for the old disk stars ({\it 25}).
Our velocity cut excludes some halo stars with lower velocities, but
permits only minimal contamination of the sample by disk stars.  By
choosing stars above 2$\sigma = 94$ km s$^{-1}$, we excluded 95\%\ of
the disk stars.

Furthermore, if a portion of our sample contains binary stars [the
fraction of white dwarfs that are in white dwarf-white dwarf pairs is
between 2 and 20\% ({\it 26})], we expect that the velocity
distribution of halo stars (Fig.\ \ref{fig:velocity}) will be biased
toward the point $(V, U) = (0, 0)$.  This is because we assumed that
all of the stars are solitary in our estimate of distance.  For those
that are actually binary, we underestimated the intrinsic luminosity
by, at most, a factor of 2 and the distance by a factor of $\sqrt{2}$.
Therefore the values of $U$ and $V$ are underestimated by the same
factor of $\sqrt{2}$.  This bias is apparent in our sample (Fig.\
\ref{fig:velocity}), implying that while we may have included a few
disk stars in our sample, we may also have excluded a few binary halo
stars.

Because we searched for stars having $\mu >
0.33^{\prime\prime}$~year$^{-1}$ there is a lack of stars with very
low space motions.  Furthermore, because of large differences in the
epochs of the plates, an object with a proper motion above
3$^{\prime\prime}$~year$^{-1}$ will have a displacement of at least
30$^{\prime\prime}$ between the plates.  This compromises the
detection of objects with extremely high space motions: even though we
searched for objects having annual motions of up to
10$^{\prime\prime}$, our survey probably had less than a 10\% chance
of finding very faint stars moving by 3$^{\prime\prime}$ year$^{-1}$
or more.  These biases against stars with very low and very high space
motions result in the dearth of points within 50 km~s$^{-1}$ of the
point $(U, V) = (0, 0)$ and the paucity of points at the extreme left
of the $UV$ velocity diagram (Fig.\ \ref{fig:velocity}).  Therefore,
we underestimated the numbers of halo white dwarfs, and we measured a
lower limit to the space density of these objects.

The subset of objects with velocities outside the 2$\sigma$ velocity
dispersion of the old disk stars (Fig.\ 3) contains 24 cool white
dwarfs and 14 white dwarfs with H$\alpha$ features (Table 1).  An
initial estimate of the space density of these objects is obtained
using only the 12 coolest white dwarfs, whose average absolute
magnitude, $M_{\rm R59F}$, is 16.  With the survey depth of 19.8 and
the survey area, a rough volume is computed yielding a space density
of $2 \times 10^{-4}$ pc$^{-3}$, or $1.2 \times 10^{-4}$ $M_{\odot}$
pc$^{-3}$.
        
To obtain a more precise estimate of the space density of all 38 of
the halo white dwarfs in our survey, we used the $1 / V_{\rm max}$
technique ({\it 27}), which determines the maximum volume, $V_{\rm
max}$, in which the survey could have found each star, given the
brightness detection limit.  The sum of $1 / V_{\rm max}$ for all of
the stars in the sample is the number density.  Because our survey was
limited to stars brighter than $R59F_{\rm lim} = 19.8$, we have
computed $V_{\rm max}$ using the absolute $R59F$ magnitudes, $M_{\rm
R59F}$, implied by the distances inferred above.

\centerline{$V_{\rm max} = (4/3) \pi d_{\rm max}^3 \Omega / (4\pi)$ pc$^{3}$ \hfil(1)}

\noindent
Here, $\Omega$ is the surveyed area in steradians, and $d_{\rm max}$,
the maximum distance in parsecs (pc) that determines $V_{\rm max}$, is
the minimum of the two following relations.

\centerline{log $d_{\rm max}{\rm (pc)} = 0.2(R59F_{\rm lim} - M_{\rm R59F})+1$ \hfil (2)}

\centerline{$d_{\rm max}$ (pc) $= d \mu / 0.33$ \hfil (3)}

The number density derived in this manner is $1.8 \times 10^{-4}$
pc$^{-3}$.  Using 0.6 $M_\odot$ as the average white dwarf mass in the
halo ({\it 28}), the local space density of halo white dwarfs is $1.1
\times 10^{-4}$ $M_\odot$ pc$^{-3}$.  However, our survey is not
complete to the $R59F = 19.8$ level.  A measure of the uniformity of
our sample is given by the average value of $V / V_{\rm max}$, where
$V$ is the volume of space given by the distance to the given star and
the survey area.  In the case of this survey, the average of $V /
V_{\rm max}$ is 0.46.  For a sample distributed uniformly in space,
this value should be 0.5.  An average of $V / V_{\rm max} = 0.5$ is
found for $R59F = 19.7$, indicating that all stars brighter than 19.7,
with the proper motion constraint mentioned above, were found in the
survey.  Recalculating the space density using $R59F_{\rm lim} = 19.7$
yields $1.3 \times 10^{-4}$ $M_\odot$ pc$^{-3}$.

The density of white dwarfs in the halo predicted from the subdwarf
star counts and a standard initial mass function is 1.3$\times
10^{-5}$ $M_\odot$ pc$^{-3}$ ({\it 29}), which is ten times smaller
than the value we calculate.  This means that star formation in the
early Galaxy must have favored higher-mass stars that would evolve
into the white dwarfs we are detecting now.  At the same time, this
early star formation must have under produced the low-mass subdwarfs
relative to the more recent star formation processes observed in the
disk ({\it 4}).

The estimated density of halo dark matter near the Sun is
approximately 8$\times 10^{-3}$ $M_\odot$ pc$^{-3}$ ({\it 30}).
Thus, the population we have detected accounts for about 2\%\ of the
local dark matter.  We treat this number as a lower limit for the
reasons discussed above and because it appears that we have not probed
the entire range of intrinsic luminosities of these objects.  Indeed,
the number density as a function of $M_{\rm R59F}$ suggests that we
are only detecting the rising power law of the luminosity function and
have not seen any indication of the expected turnover due to the
finite life time of the halo white dwarf population ({\it 31}).  This
means that there may be an undetected, larger population of even
fainter and cooler white dwarfs in the halo.

\smallskip

\centerline{\bf Discussion}

Microlensing experiments have indirectly revealed a population of
massive, compact objects in the line of sight to the Large Magellanic
Cloud.  The Massive Compact Halo Object project (MACHO) ({\it 1})
estimates that these compact objects contribute between 8 and 50\%\ of
the local halo dark matter at the 95\%\ confidence level, and the
inferred lens mass suggests that they are white dwarfs.  The similar
Exp\'erience pour la Recherche d'Objets Sombres project (EROS) places
a strong upper limit of 35\%\ for such objects ({\it 32}).  Because
the halo white dwarf space density that we derive from our survey
represents a lower limit and is consistent with the two measurements
mentioned above, the population of white dwarfs with halo kinematics
that we have discovered provides a natural explanation for the
microlensing results.

\begin{figure}

\centerline{\psfig{file=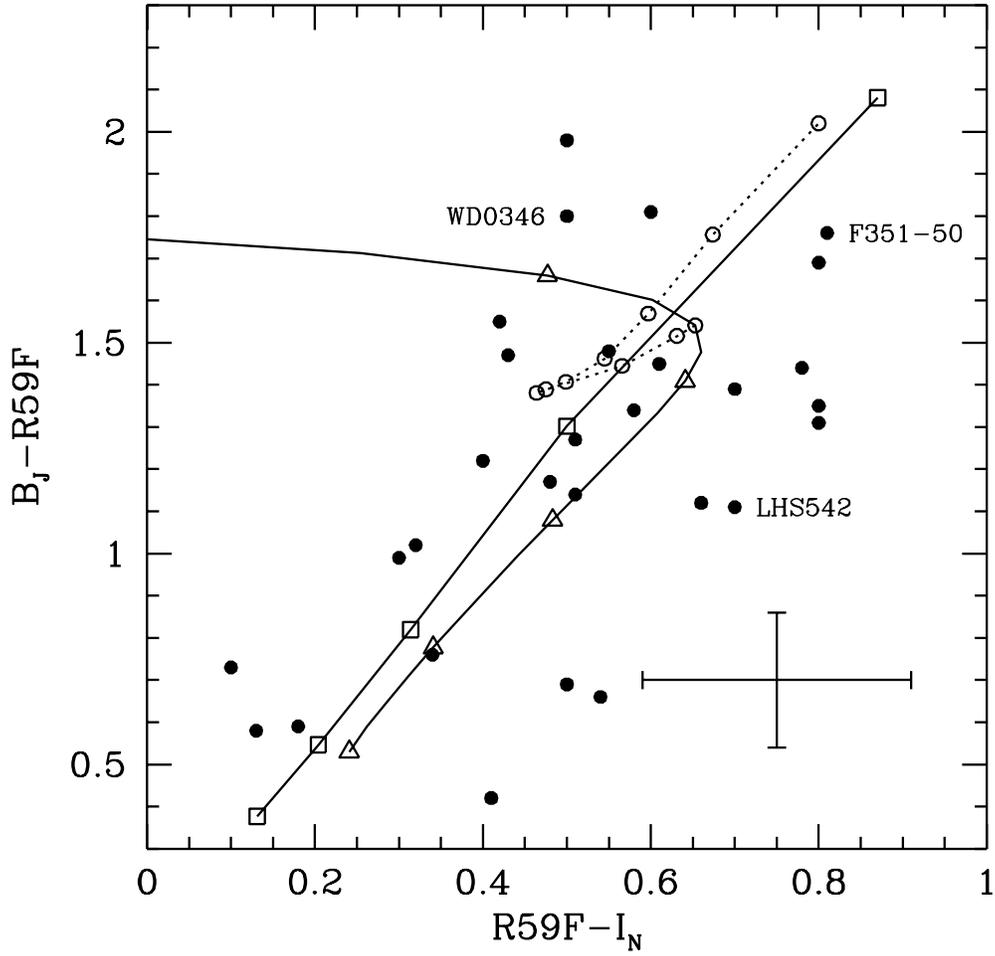,height=5.0in}}

\caption{Photographic color-color diagram.  The cooler stars listed in
Table 1 are shown as solid dots, and a typical error bar is indicated.
The synthetic colors of pure He atmospheres (open squares) ({\it 9})
form a diagonal sequence from $\Teff=8000\,$K (lower left) to
4000$\,$K, in steps of 1000$\,$K.  For pure H atmospheres (open
triangles) ({\it 9}) the sequence starts at 7000$\,$K (lower left) and
turns over to bluer $R59F - I_{\rm N}$ below 3500$\,$K due to the
onset of H$_2$ collision-induced opacity. The dotted line connecting
open circles shows the effect of mixed H/He composition on a
$\Teff=3500\,$K model, with the number ratio of He to H ranging from 0
to $10^7$ ({\it 7}). All models have $\log g $(cm s$^{-2}$)$=8$.  As a
basis for comparison, three stars whose spectra have been studied in
greater detail ({\it 3, 7}) are labeled.  WD0346$+$246 is not part of
our survey and is shown here for comparison.
\label{fig:color}}
\end{figure}

A number of the white dwarfs revealed in this survey may be cooler
than any other white dwarf previously known.  There are at least three
objects (Table 1) that are comparable in temperature to WD0346$+$246
({\it 7}), representing a doubling of the number of ultracool white
dwarfs known.  The nature of cool white dwarfs can be discussed on the
basis of their positions in color-color diagrams.  Although a
combination of optical and infrared colors is necessary at these low
effective temperatures ({\it 7, 23}), for now, we must contend with the
limited spectral coverage of photographic plates for this sample.  The
comparison of $B_{\rm J}$, $R59F$, and $I_{\rm N}$ (a bandpass centered
on 0.8 $\mu$m) colors is a poor diagnostic of atmospheric composition
for effective temperatures, $\Teff$, greater than 3500 K (Fig.\
\ref{fig:color}).  Given the large photometric error bars, there is
little that distinguishes the colors of these stars from a sample of
cool white dwarfs with disk kinematics ({\it 33}).

Stars with $B_{\rm J} - R59F \wig< 1.2$ are relatively warm, with
$\Teff \wig> 5000\,$K (except for the strange stars LHS 3250, SDSS
1337$+$00 and LHS 1402).  They are also the most luminous and most
distant stars in the sample (Table 1).  Although it is not possible to
guess the atmospheric composition of any of these stars from the
present data, we point out that the sample may contain a few unusual
objects, in addition to the peculiar LHS 1402 and WD2356$-$209 (Fig.\
2b).  Other inferences, however, can be made from the photographic
color-color diagram (Fig.\ \ref{fig:color}), using stars previously
studied.  LHS 542 has a pure He atmosphere and $\Teff = 4747\,$K ({\it
24}). WD0346$+$246 and F351$-$50 are similar to each other with
predominantly He atmospheres, a very small admixture of H, and $\Teff$
of $\sim 3750\,$K and $\sim 3500\,$K, respectively ({\it 7}).  The
most extreme star of the sample, and also the faintest, is
WD0351$-$564 with $B_{\rm J} - R59F = 1.98$.  The stars with $B_{\rm
J} - R59F > 1.6$ (WD0351$-$564, WD0205$-$053 and WD0345$-$362) may
have either pure He atmospheres with $\Teff > 4000\,$K or, more
interestingly, be other examples of very cool white dwarfs with He
atmospheres polluted by a small amount of H, like WD0346$+$246.  The
stars most likely to have pure H atmospheres residing below the color
turnoff due to strong H$_2$ collision-induced opacity are
WD0135$-$546, WD0340$-$330, and, perhaps, WD0351$-$564.

We still do not understand the nature of the three strange objects LHS
3250, LHS 1402, and SDSS 1337$+$00 ({\it 5-7}), all of which lie at
$R59F - I_{\rm N}$ colors near $-$0.5 (well outside of Fig.\
\ref{fig:color}).  The kinematics of these stars, with the exception
of LHS 1402, suggest that they may be disk objects.  Although they are
certainly cool white dwarfs, we know neither their effective
temperatures nor their atmospheric compositions.  They may represent
another, perhaps unusual, stage in the spectral evolution of cool
white dwarfs.

The direct detection of a significant white dwarf halo population
opens promising avenues of investigation.  Besides the opportunity to
study individual white dwarfs that are older and cooler than any known
so far, dedicated studies of this population of stellar remnants will
bear directly on the ancient history of the Galaxy.  For instance, a
determination of the luminosity function of halo white dwarfs and a
characterization of the low luminosity cutoff will reveal the age of
the first generation of halo stars, lead to a determination of its
initial mass function, and provide strong constraints on formation
scenarios of the Galaxy.  The interpretation of observations of young
galaxies at high redshifts will benefit from a better understanding of
the early history of our Galaxy.  Finally, the complete
characterization of this new stellar population will reveal whether it
is responsible for the microlensing.


\renewcommand{\baselinestretch}{1.07}

\centerline{\bf References}

\noindent
1. C.~Alcock {\it et al.}, {\it Astrophys.\ J.} {\bf 542}, 281 (2000).

\noindent
2. S.~T.~Hodgkin {\it et al.}, {\it Nature} {\bf 403}, 57 (2000).

\noindent
3. R.~Ibata, M.~J.~Irwin, O.~Bienaym\'e, R.~Scholz, J.~Guibert, {\it
   Astrophys.\ J.\ Lett.} {\bf 532}, L41 (2000).

\noindent
4. G.~Chabrier, {\it Astrophys.\ J.\ Lett.} {\bf 513}, L103 (1999).

\noindent
5. H.~C.~Harris {\it et al.}, {\it Astrophys.\ J.} {\bf 524}, 1000 (1999).

\noindent
6. H.~C.~Harris {\it et al.}, {\it Astrophys.\ J. Lett.}, {\bf 549}, L109 (2001).

\noindent
7. B.~R.~Oppenheimer {\it et al.}, {\it Astrophys.\ J.}, {\bf 550}, 448 (2001).

\noindent
8. B.~Hansen, {\it Nature} {\bf 394}, 860 (1998). 

\noindent
9. P.~Bergeron, D.~Saumon, F.~Wesemael, {\it Astrophys.\ J.} {\bf
   443}, 764 (1995), and D.~Saumon, S.~Jacobson, {\it Astrophys.\ J.\
   Lett.} {\bf 511}, L107 (1999).

\noindent
10. B.~Hansen, {\it Astrophys.\ J.} {\bf 520}, 680 (1999).

\noindent
11. A.~Borysow, L.~Frommhold, {\it Astrophys.\ J.\ Lett.} {\bf 348},
    L41 (1990).

\noindent
12. N.~C.~Hambly {\it et al.}, in preparation (2001).

\noindent
13. N.~C.~Hambly, M.~J.~Irwin, H.~T.~MacGillivray, in preparation (2001).

\noindent
14. N.~C.~Hambly, A.~C.~Davenhall, M.~J.~Irwin, H.~T.~MacGillivray, in
    preparation (2001).

\noindent
15. N.~C.~Hambly, in {\it The 12th European Conference on White
    Dwarfs}, Newark, Delaware, 12 to 16 July 2000, H. Shipman and
    J. Provencal, Eds. (Astronomical Society of the Pacific, San
    Francisco, 2001).

\noindent
16. D.~G.~Monet {\it et al.}, {\it Astron.\ J.} {\bf 120}, 1541 (2000).

\noindent
17. W.~J.~Luyten, {\it NLTT Catalog} (Univ. Minnesota Press,
    Minneapolis, 1979).

\noindent
18. R.~A.~Knox, M.~R.~S.~Hawkins, N.~C.~Hambly, {\it Mon.\ Not.\ Royal
    Astron.\ Soc.} {\bf 306}, 736 (1999).

\noindent
19. C.~Flynn, J.~Sommer--Larsen, B.~Fuchs, D.~S.~Graff, {\it Mon.\
    Not.\ Royal Astron.\ Soc.}, {\bf 322}, 553 (2001).

\noindent
20. R.~D.~Scholz, M.~J.~Irwin, R.~Ibata, H.~Jahrei\ss, O.~Yu.~Malkov,
    {\it Astron.\ Astrophys.} {\bf 353}, 958 (2000).

\noindent
21. N.~C.~Hambly, S.~J.~Smartt, S.~T.~Hodgkin, {\it Astrophys.\ J.\
    Lett.} {\bf 489}, L157 (1997).

\noindent
22. N.~C.~Hambly {\it et al.}, {\it Mon.\ Not.\ Royal Astron.\ Soc.}
    {\bf 309}, L33 (1999).

\noindent
23. P.~Bergeron, M.-T.~Ruiz, S.~K.~Leggett, {\it Astrophys.\ J.\
    Suppl.\ Ser.} {\bf 108}, 339 (1997).

\noindent
24. S.~K.~Leggett, M.-T.~Ruiz, P.~Bergeron, {\it Astrophys.\ J.\
    Lett.} {\bf 497}, L294 (1998).

\noindent
25. M.~Chiba, T.~C.~Beers, {\it Astron.\ J.} {\bf 119}, 2843 (2000).

\noindent
26. P.~F.~L.~Maxted, P.~R.~Marsh, {\it Mon.\ Not.\ Royal Astron.\ Soc.}
    {\bf 307}, 122 (1999).

\noindent
27. M.~A.~Wood, T.~D.~Oswalt, {\it Astrophys.\ J.} {\bf 497}, 870 (1998).

\noindent
28. P.~Bergeron, R.~A.~Saffer, J.~Liebert, {\it Astrophys.\ J.} {\bf
    394}, 228 (1992).

\noindent
29. A.~Gould, C.~Flynn, J.~N.~Bahcall, {\it Astrophys.\ J.} {\bf
    503}, 798 (1998).

\noindent
30. E.~I.~Gates, G.~Gyuk, M.~S.~Turner, {\it Astrophys.\ J.\ Lett.}
    {\bf 449}, L123 (1995).

\noindent
31. G.~Fontaine, P.~Brassard, P.~Bergeron, {\it Publ.\ Astron.\ Soc.\
    Pac.}, {\bf 113}, 409 (2001). See also C.~M.~Tamanaha, J.~Silk,
    M.~A.~Wood, D.~E.~Winget, {\it Astrophys.\ J.} {\bf 358}, 164
    (1990).

\noindent
32. T.~Lasserre {\it et al.}, {\it Astron.\ Astrophys.} {\bf 355},
    L39 (2000).

\noindent
33. P.~Bergeron, S.~K.~Leggett, M.-T.~Ruiz, {\it Astrophys.\ J.}, in press.

\noindent
34. We thank M. Irwin, I. King, G. Chabrier and T.  Nakajima for
    helpful discussions.  B.R.O. is supported by a Hubble Postdoctoral
    Research Fellowship, grant number HF-01122.01-99A.  D.S.
    acknowledges support from NSF Grant AST97-31438.

Note added after Science Express online publication: This text
reflects the correction of a few typographical errors in the online
version of the table.  It also includes the new constraint on the
calculation of $d_{\rm max}$ which accounts for the fact that the
survey could not have detected stars with proper motions below 0.33
arcseconds per year.

16 February 2001; accepted 13 March 2001  

Published online 22 March 2001; 

10.1126/science.1059954  

Include this information when citing this paper.

\clearpage

\begin{sidewaystable}

{
\renewcommand{\baselinestretch}{1.0}
\renewcommand{\arraystretch}{.8}
\small
\begin{center}

Table 1. Candidate Halo White Dwarfs.

\begin{tabular}{llrccccccc}

\hline

Star & \multicolumn{2}{c}{R.~A.~and Dec.} & $\mu$ & PA  & $B_{\rm J}$ & ($B_{\rm J}- R59F$) & 
$(R59F - I_{\rm N})$ & $v_{\rm tan}$(est) & Dist. \\
Name & \multicolumn{2}{c}{(Equinox \& Epoch J2000)} & $^{\prime\prime}$ year$^{-1}$ & 
$^{\circ}$ & (mag) & (mag) & (mag) & (km s$^{-1}$) & (pc)\\

\hline

F351--50 &     00:45:19.695 & $-33$:29:29.46 & $ 2.490\pm0.069$ &  128.0 & $20.13$ & $ 1.76$ & $ 0.81$ &    440 & 37 \\
WD0153--014$^{c}$ &  01:53:51.454 & $-01$:23:40.76 & $ 0.395\pm0.010$ &  170.8 & $18.87$ & $ 0.15$ & 0.0$^{b}$ & 265 & 141 \\ 
LHS 542 &      23:19:09.518 & $-06$:12:49.92 & $ 1.695\pm0.029$ &  200.7 & $18.72$ & $ 1.11$ & $ 0.70$ &    339 & 42  \\ 
WD0351--564 &  03:51:09.376 & $-56$:27:07.12 & $ 1.085\pm0.027$ &  165.8 & $21.70$ & $ 1.98$ & 0.5$^{b}$ &  305 & 59  \\ 
LHS 147$^{c}$ &  01:48:09.120 & $-17$:12:14.08 & $ 1.095\pm0.028$ &  186.6 & $17.96$ & $ 0.38$ & $ 0.30$ &  367 & 71 \\ 
WD2326--272 &  23:26:10.718 & $-27$:14:46.68 & $ 0.603\pm0.057$ &   99.5 & $20.46$ & $ 0.99$ & 0.3$^{b}$ &  310 & 108 \\ 
WD0135--039$^{c}$ &  01:35:33.685 & $-03$:57:17.90 & $ 0.503\pm0.020$ &  111.0 & $19.92$ & $ 0.53$ & $ 0.34$ & 348 & 146 \\
LHS 4042$^{c}$    &  23:54:35.034 & $-32$:21:19.44 & $ 0.427\pm0.020$ &   95.7 & $17.66$ & $ 0.11$ & $ 0.24$ & 172 & 85 \\ 
WD2356--209 &  23:56:45.091 & $-20$:54:49.31 & $ 0.391\pm0.044$ &  237.3 & $21.12$ & $ 1.45$ & $ 0.61$ &    158 & 85 \\ 
WD0227--444 &  02:27:29.562 & $-44$:23:08.64 & $ 0.344\pm0.021$ &  129.0 & $20.40$ & $ 1.27$ & $ 0.51$ &    124 & 76 \\ 
J0014--3937 &  00:13:47.465 & $-39$:37:24.00 & $ 0.739\pm0.024$ &  198.5 & $19.35$ & $ 1.39$ & $ 0.70$ &    140 & 40  \\ 
LHS 4033$^{c}$    &  23:52:31.941 & $-02$:53:11.76 & $ 0.723\pm0.056$ &   62.8 & $17.40$ & $ 0.26$ & $ 0.05$ & 216 & 63 \\ 
LP 586--51$^{c}$ &  01:02:07.181 & $-00$:33:01.82 & $ 0.417\pm0.105$ &  128.7 & $18.44$ & $ 0.12$ & $-0.16$\phantom{$-$} & 238 & 120 \\ 
WD2242--197 &  22:41:44.252 & $-19$:40:41.41 & $ 0.367\pm0.040$ &   76.5 & $20.04$ & $ 0.76$ & $ 0.34$ &    204 & 117 \\ 
WD0205--053 &  02:05:11.620 & $-05$:17:54.33 & $ 1.051\pm0.025$ &   67.6 & $19.57$ & $ 1.69$ & $ 0.80$ &    156 & 31 \\ 
WD0100-645$^{c}$ &   01:00:50.394 & $-64$:29:11.21 & $ 0.560\pm0.048$ &   70.9 & $17.98$ & $ 0.70$ & $ 0.30$ & 130 & 49 \\ 
WD0125--043 &  01:25:05.884 & $-04$:17:02.56 & $ 0.409\pm0.135$ &   80.6 & $20.26$ & $ 1.02$ & $ 0.32$ &    185 & 95 \\ 
WD2346--478$^{c}$ &  23:46:02.857 & $-47$:51:01.92 & $ 0.533\pm0.050$ &  219.6 & $18.44$ & $ 0.99$ & $ 0.42$ & 108 & 43 \\ 
LHS 1447 &     02:48:13.182 & $-30$:01:32.40 & $ 0.550\pm0.055$ &   51.4 & $18.78$ & $ 0.59$ & $ 0.18$ &    210 & 80 \\ 
WD0300--044 &  03:00:23.644 & $-04$:25:24.78 & $ 0.390\pm0.026$ &  135.8 & $20.15$ & $ 0.73$ & 0.1$^{b}$ &  237 & 128 \\
WD0123--278 &  01:23:03.784 & $-27$:48:14.59 & $ 0.352\pm0.040$ &   68.4 & $20.95$ & $ 1.35$ & 0.8$^{b}$ &  148 & 89 \\ 
WD2259--465 &  22:59:06.633 & $-46$:27:58.86 & $ 0.414\pm0.032$ &  112.4 & $20.37$ & $ 1.31$ & 0.8$^{b}$ &  140 & 71 \\ 
WD0340--330 &  03:40:08.673 & $-33$:01:00.30 & $ 0.606\pm0.069$ &  128.0 & $20.56$ & $ 1.55$ & $ 0.42$ &    168 & 58 \\ 
LHS 1402$^{a}$ & 02:24:32.255 & $-28$:54:59.36 & $0.490\pm0.030$ &  92.6 & $18.32$ & $ 0.46$ & $-0.63$\phantom{$-$} &  176 & 76 \\
LHS 1274$^{c}$ &  01:39:14.380 & $-33$:49:03.31 & $ 0.579\pm0.037$ &   93.4 & $17.71$ & $ 0.52$ & $ 0.34$ &  147 & 53 \\ 
WD0214--419 &  02:14:14.887 & $-41$:51:09.04 & $ 0.334\pm0.026$ &  106.7 & $20.63$ & $ 1.17$ & $ 0.48$ &    150 & 95 \\ 
WD0044--284 &  00:44:02.188 & $-28$:24:11.15 & $ 0.348\pm0.063$ &  191.1 & $20.83$ & $ 1.48$ & $ 0.55$ &    119 & 72 \\ 
WD2214--390$^{c}$ &  22:14:34.727 & $-38$:59:07.05 & $ 1.056\pm0.013$ &  110.1 & $16.57$ & $ 0.75$ & $ 0.40$ &  120 & 24 \\ 
WD2324--595$^{c}$ &  23:24:10.165 & $-59$:28:07.95 & $ 0.581\pm0.016$ &  167.1 & $17.13$ & $ 0.23$ & $ 0.05$ &  159 & 58 \\ 
LP 588--37$^{c}$ &  01:42:20.770 & $-01$:23:51.38 & $ 0.354\pm0.010$ &  162.8 & $18.62$ & $ 0.19$ & 0.0$^{b}$ & 202 & 120 \\ 
WD0345--362 &  03:45:32.711 & $-36$:11:03.60 & $ 0.589\pm0.081$ &  167.8 & $21.12$ & $ 1.81$ & $ 0.60$ &    155 & 55 \\ 
WD0045--061 &  00:45:06.325 & $-06$:08:19.65 & $ 0.699\pm0.017$ &  171.2 & $18.82$ & $ 1.12$ & $ 0.66$ &    145 & 44 \\ 
WD0225--326 &  02:25:28.681 & $-32$:37:53.92 & $ 0.338\pm0.033$ &   59.7 & $18.94$ & $ 0.58$ & $ 0.13$ &    140 & 88 \\ 
WD2348--548 &  23:48:46.904 & $-54$:45:46.21 & $ 0.388\pm0.033$ &  105.5 & $19.76$ & $ 1.14$ & $ 0.51$ &    121 & 66 \\ 
WD0117--268 &  01:17:51.649 & $-26$:48:51.21 & $ 0.468\pm0.032$ &   82.2 & $19.91$ & $ 1.34$ & $ 0.58$ &    122 & 55 \\ 
LP 651--74$^{c}$ &  03:07:14.113 & $-07$:14:59.12 & $ 0.477\pm0.036$ &  203.8 & $17.56$ & $ 0.94$ & $ 0.4^{b}$ & 68 & 30 \\ 
WD0135--546 &  01:35:38.677 & $-54$:35:27.78 & $ 0.675\pm0.018$ &   80.9 & $19.52$ & $ 1.47$ & $ 0.43$ &  127 & 39 \\ 
WD0100--567$^{c}$ &  01:00:43.076 & $-56$:46:36.61 & $ 0.414\pm0.043$ &  45.1 & $17.85$ & $ 0.63$ & $ 0.4^{b}$ & 98 & 50 \\ 

\hline

\end{tabular}

\end{center}

$^a$This object is extremely cool but has a spectrum similar to the
peculiar LHS 3250 ({\it 7}).

$^b$These R59F $-$ I$_{\rm N}$ colors are calculated from
spectrophotometry.

$^c$These stars have H$\alpha$ features.  All others have featureless
spectra.

}
\end{sidewaystable}

\begin{table}
\caption{Candidate Halo White Dwarfs. WD names are new discoveries;
otherwise, Luyten LHS or LP names are given; J0014--3937 was discovered
by Scholz {\it et al.} ({\it 20}), who classify this object as ``cool
no H$\alpha$'', and F351--50 was discovered by Ibata {\it et al.}
({\it 3}).  The tangential velocity, $v_{\rm tan}$(est) in km
s$^{-1}$, is an estimate based on photometric parallax.  Distances are
estimates based on a color-magnitude relation (see text).  Star
positions are given in right ascension (RA) and declination (Dec.) at the equinox
and epoch of J2000.  The proper motion, $\mu$, and position angle, PA,
are given in $^{\prime\prime}$~year$^{-1}$ and degrees,
respectively. The photometric measurements in $B_{\rm J}$, $R59F$, and
$I_{\rm N}$ are given in magnitudes, with uncertainties of $\pm 0.1$
mag.\ on average.  The distance is in parsecs from the Sun.}
\end{table}

\clearpage
\renewcommand{\baselinestretch}{1.7}

\clearpage 

\clearpage

\clearpage 

\end{document}